# Ten Research Challenge Areas in Data Science

Jeannette M. Wing
Avanessians Director of the Data Science Institute and Professor of Computer Science
Columbia University

December 30, 2019

Although data science builds on knowledge from computer science, mathematics, statistics, and other disciplines, data science is a unique field with many mysteries to unlock: challenging scientific questions and pressing questions of societal importance.

**Is data science a discipline?**

Data science is a field of study: one can get a degree in data science, get a job as a data scientist, and get funded to do data science research.  But is data science a discipline, or will it evolve to be one, distinct from other disciplines?  Here are a few meta-questions about data science as a discipline.

- *What is/are the driving deep question(s) of data science?*  Each scientific discipline (usually) has one or more "deep" questions that drive its research agenda: What is the origin of the universe (astrophysics)?  What is the origin of life (biology)?  What is computable (computer science)?  Does data science inherit its deep questions from all its constituency disciplines or does it have its own unique ones?

- *What is the role of the domain in the field of data science?*  People (including this author) (Wing, Janeia, Kloefkorn, & Erickson 2018) have argued that data science is unique in that it is not just about methods, but about the use of those methods in the context of a domain—the domain of the data being collected and analyzed; the domain for which a question to be answered comes from collecting and analyzing the data.  Is the inclusion of a domain inherent in defining the field of data science?  If so, is the way it is included unique to data science?

- *What makes data science data science?*  Is there a problem unique to data science that one can convincingly argue would not be addressed or asked by any of its constituent disciplines, e.g., computer science and statistics?

**Ten research areas**

While answering the above meta-questions is still under lively debate, we can ask an easier question, one that also underlies any field of study: What are the research challenge areas that drive the study of data science?  Here is a list of ten.  They are not in any priority order, and some of them are related to each other.  They are phrased as challenge areas, not challenge questions.  They are not necessarily the "top ten" but they are a good ten to start the community discussing what a broad research agenda for data science might look like.[1]

1. *Scientific understanding of learning, especially deep learning algorithms.* As much as we admire the astonishing successes of deep learning, we still lack a scientific understanding of why deep learning works so well. We do not understand the mathematical properties of deep learning models. We do not know how to explain why a deep learning model produces one result and not another. We do not understand how robust or fragile they are to perturbations to input data distributions. We do not understand how to verify that deep learning will perform the intended task well on new input data. Deep learning is an example of where experimentation in a field is far ahead of any kind of theoretical understanding.

2. *Causal reasoning.* Machine learning is a powerful tool to find patterns and examine correlations, particularly in large data sets. While the adoption of machine learning has opened many fruitful areas of research in economics, social science, and medicine, these fields require methods that move beyond correlational analyses and can tackle causal questions. A rich and growing area of current study is revisiting causal inference in the presence of large amounts of data. Economists are already revisiting causal reasoning by devising new methods at the intersection of economics and machine learning that make causal inference estimation more efficient and flexible (Athey, 2016), (Taddy, 2019). Data scientists are just beginning to explore multiple causal inference, not just to overcome some of the strong assumptions of univariate causal inference, but because most real-world observations are due to multiple factors that interact with each other (Wang & Blei, 2018).

3. *Precious data.* Data can be precious for one of three reasons: the dataset is expensive to collect; the dataset contains a rare event (low signal-to-noise ratio*);* or the dataset is artisanal—small and task-specific. A good example of expensive data comes from large, one-of, expensive scientific instruments, e.g., the Large Synoptic Survey Telescope, the Large Hadron Collider, the IceCube Neutrino Detector at the South Pole. A good example of rare event data is data from sensors on physical infrastructure, such as bridges and tunnels; sensors produce a lot of raw data, but the disastrous event they are used to predict is (thankfully) rare. Rare data can also be expensive to collect. A good example of artisanal data is the tens of millions of court judgments that China has released online to the public since 2014 (Liebman, Roberts, Stern, & Wang, 2017) or the 2+ million US government declassified documents collected by Columbia's History Lab (Connelly, Madigan, Jervis, Spirling, & Hicks, 2019). For each of these different kinds of precious data, we need new data science methods and algorithms, taking into consideration the domain and intended uses of the data.

4. *Multiple, heterogeneous data sources.* For some problems, we can collect lots of data from different data sources to improve our models. For example, to predict the effectiveness of a specific cancer treatment for a human, we might build a model based on 2-D cell lines from mice, more expensive 3-D cell lines from mice, and the costly DNA sequence of the cancer cells extracted from the human. State-of-the-art data science methods cannot as yet handle combining multiple, heterogeneous sources of data to build a single, accurate model.  Since many of these data sources might be precious data, this challenge is related to the third challenge.  Focused research in combining multiple sources of data will provide extraordinary impact.

---

[1] Based on an NSF workshop, the recent report by the statistics community, "Statistics at a Crossroads: Who is for the Challenge?," gives an overlapping list of seven foundational challenges in data science (Berger et al., 2019).

5. *Inferring from noisy and/or incomplete data.*  The real world is messy and we often do not have complete information about every data point.  Yet, data scientists want to build models from such data to do prediction and inference.  A great example of a novel formulation of this problem is the planned use of differential privacy for Census 2020 data (Garfinkel, 2019), where noise is deliberately added to a query result, to maintain the privacy of individuals participating in the census. Handling "deliberate" noise is particularly important for researchers working with small geographic areas such as census blocks, since the added noise can make the data uninformative at those levels of aggregation. How then can social scientists, who for decades have been drawing inferences from census data, make inferences on this "noisy" data and how do they combine their past inferences with these new ones? Machine learning's ability to better separate noise from signal can improve the efficiency and accuracy of those inferences.

6. *Trustworthy AI.*  We have seen rapid deployment systems using artificial intelligence (AI) and machine learning in critical domains such as autonomous vehicles, criminal justice, healthcare, hiring, housing, human resource management, law enforcement, and public safety, where decisions taken by AI agents directly impact human lives. Consequently, there is an increasing concern if these decisions can be trusted to be correct, reliable, robust, safe, secure, and fair, especially under adversarial attacks. One approach to building trust is through providing explanations of the outcomes of a machine learned model.  If we can interpret the outcome in a meaningful way, then the end user can better trust the model.  Another approach is through formal methods, where one strives to prove once and for all a model satisfies a certain property. New trust properties yield new tradeoffs for machine learned models, e.g., privacy versus accuracy; robustness versus efficiency. There are actually multiple audiences for trustworthy models: the model developer, the model user, and the model customer.  Ultimately, for widespread adoption of the technology, it is the public who must trust these automated decision systems.

7. *Computing systems for data-intensive applications.*  Traditional designs of computing systems have focused on computational speed and power: the more cycles, the faster the application can run.  Today, the primary focus of applications, especially in the sciences (e.g., astronomy, biology, climate science, materials science), is data.  Also, novel special-purpose processors, e.g., GPUs, FPGAs, TPUs, are now commonly found in large data centers. Even with all these data and all this fast and flexible computational power, it can still take weeks to build accurate predictive models; however, applications, whether from science or industry, want *real-time* predictions.  Also, data-hungry and compute-hungry algorithms, e.g., deep learning, are energy hogs (Strubell, Ganesh, & McCallum, 2019).  Not only should we consider space and time, but energy consumption, in our performance metrics.  In short, we need to rethink computer systems design from first principles, with data (not compute) the focus.  New computing systems designs need to consider: heterogeneous processing; efficient layout of massive amounts of data for fast access; the target domain, application, or even task; and energy efficiency.

8. *Automating front-end stages of the data life cycle.* While the excitement in data science is due largely to the successes of machine learning, and more specifically deep learning, before we get to use machine learning methods, we need to prepare the data for analysis. The early stages in the data life cycle (Wing, 2019) are still labor intensive and tedious. Data scientists, drawing on both computational and statistical methods, need to devise automated methods that address data cleaning and data wrangling, without losing other desired properties, e.g., accuracy, precision, and robustness, of the end model. One example of emerging work in this area is the Data Analysis Baseline Library (Mueller, 2019), which provides a framework to simplify and automate data cleaning, visualization, model building, and model interpretation. The Snorkel project addresses the tedious task of data labeling (Ratner et al., 2018).

9. *Privacy.* Today, the more data we have, the better the model we can build. One way to get more data is to share data, e.g., multiple parties pool their individual datasets to build collectively a better model than any one party can build. However, in many cases, due to regulation or privacy concerns, we need to preserve the confidentiality of each party's dataset. An example of this scenario is in building a model to predict whether someone has a disease or not. If multiple hospitals could share their patient records, we could build a better predictive model; but due to Health Insurance Portability and Accountability Act (HIPAA) privacy regulations, hospitals cannot share these records. We are only now exploring practical and scalable ways, using cryptographic and statistical methods, for multiple parties to share data and/or share models to preserve the privacy of each party's dataset. Industry and government are exploring and exploiting methods and concepts, such as secure multi-party computation, homomorphic encryption, zero-knowledge proofs, and differential privacy, as part of a point solution to a point problem.

10. *Ethics.* Data science raises new ethical issues. They can be framed along three axes: (1) the ethics of data: how data are generated, recorded, and shared; (2) the ethics of algorithms: how artificial intelligence, machine learning, and robots interpret data; and (3) the ethics of practices: devising responsible innovation and professional codes to guide this emerging science (Floridi & Taddeo, 2016) and for defining Institutional Review Board (IRB) criteria and processes specific for data (Wing, Janeia, Kloefkorn, & Erickson 2018). Example ethical questions include how to detect and eliminate racial, gender, socio-economic, or other biases in machine learning models.

**Closing remarks**

As many universities and colleges are creating new data science schools, institutes, centers, etc. (Wing, Janeia, Kloefkorn, & Erickson 2018), it is worth reflecting on data science as a field. Will data science as an area of research and education evolve into being its own discipline or be a field that cuts across all other disciplines? One could argue that computer science, mathematics, and statistics share this commonality: they are each their own discipline, but they each can be applied to (almost) every other discipline. What will data science be in 10 or 50 years?

**Acknowledgments**

I would like to thank Cliff Stein, Gerad Torats-Espinosa, Max Topaz, and Richard Witten for their feedback on earlier renditions of this paper. Many thanks to all Columbia Data Science faculty who have helped me formulate and discuss these tend (and other) challenges during our Fall 2019 DSI retreat.